\documentclass[twocolumn]{article}

\usepackage[a4paper,margin=1in]{geometry}

\usepackage{amsthm}
\usepackage{amsmath}
\usepackage{amssymb}
\usepackage{dsfont}
\usepackage{mathtools}
\usepackage[notrig]{physics}
\usepackage{multirow}
\usepackage{tikz}
\usepackage{quantikz}
\usepackage{authblk}
\usepackage{xcolor}
\usepackage{hyperref}
\usepackage{subcaption}
\usepackage[capitalise]{cleveref}
\usepackage{algcompatible}
\usepackage{algorithm}
\floatname{algorithm}{Protocol}
\crefname{algorithm}{Protocol}{Protocols}
\usepackage [
  n,
  advantage,
  operators,
  sets,
  adversary,
  landau,
  probability, 
  notions,
  logic,
  ff,
  mm,
  primitives,
  events,
  complexity,
  asymptotics,
  keys
  ] {cryptocode}

\usepackage[style=numeric-comp,sorting=none]{biblatex}
\usetikzlibrary{decorations.pathreplacing,calligraphy}
\newcommand{\Redo}{\mathsf{Redo}}
\newcommand{\Abort}{\mathsf{Abort}}
\newcommand{\Ok}{\mathsf{Ok}}

\newtheorem{theorem}{Theorem}

\makeatletter
\DeclareRobustCommand{\rvdots}{%
  \vbox{
    \baselineskip4\p@\lineskiplimit\z@
    \kern-\p@
    \hbox{.}\hbox{.}\hbox{.}
  }}
\makeatother

\title{On-Chip Verified Quantum Computation with an Ion-Trap Quantum Processing Unit}

\author[1,*]{Cica Gustiani}
\author[2,*]{Dominik Leichtle}
\author[3,*]{Daniel Mills}
\author[2]{Jonathan Miller}
\author[2]{Ross Grassie}
\author[1,2]{Elham Kashefi}

\affil[1]{\small \textit{LIP6, CNRS, Sorbonne Université, 4 Place Jussieu, 75005 Paris, France}}
\affil[2]{\small \textit{School of Informatics, University of Edinburgh, 10 Crichton Street, EH8 9AB Edinburgh, United Kingdom}}
\affil[3]{\small \textit{Quantinuum, Terrington House, 13-15 Hills Road, Cambridge CB2 1NL, UK}}
\affil[*]{Corresponding authors: \href{mailto:cicagustiani@gmail.com}{cica.gustiani@lip6.fr}, \href{mailto:dominik.leichtle@ed.ac.uk}{dominik.leichtle@ed.ac.uk}, \href{mailto:daniel.mills@quantinuum.com}{daniel.mills@quantinuum.com}}

\date{}

\begin{document}

\maketitle

\begin{abstract}

    \bfseries
    We present and experimentally demonstrate a novel approach to verifying and benchmarking quantum computing, implementing it on an ion-trap quantum computer. Unlike previous information-theoretically secure verification protocols, which typically require quantum communication between Client and Server, our approach is implemented entirely on-chip. This eliminates the need for a quantum client and significantly enhances practicality.
    
    We perform tomography to justify the additionally required assumption that the noise is independent of the secret used to prepare the Server's single-qubit states. We quantify the soundness error which may be caused by residual secret dependencies. We demonstrate our protocol on the 20-qubit Quantinuum H1-1 ion-trap quantum processing unit, using qubit measurements and resets to construct measurement patterns with up to 52 vertices. To our knowledge, these are the largest verified measurement-based quantum computations performed to date.
    
    Our results pave the way for more accessible and efficient verification and benchmarking strategies in near-term quantum devices, enabling robust performance assessment without the added cost of external quantum infrastructure.

\end{abstract}

\section{Introduction}

As quantum technology matures it brings with it a diverse array of exciting but difficult to classically simulate applications. While present and near future Quantum Processing Units (QPUs) remain delicate and susceptible to noise, scalable and rigorous methods for verification will be essential to ensure confidence in the output of the computations.

On commercially available QPUs, the most commonly used techniques broadly measure the device's susceptibility to noise in order to gauge the accuracy of the outcome of a computation. A selection of benchmarking techniques exist to provide measures of the accuracy one can expect from a QPU \cite{Eisert_2020}. These may either be indicative of component \cite{Helsen_2022}, or holistic performance \cite{Mills_2021, lubinski2024quantumalgorithmexplorationusing, Cross_2019}, from which the performance of a particular computation can be inferred. However this inference requires either that some specific noise model is present, that the noise is separable between qubits, or that the noise depends only on some general characteristics of the circuit (such as two qubit gate count). When these assumptions are not met such schemes give no clear guarantees for the correct performance of arbitrary quantum algorithms.
Aerifiable Universal Blind Quantum Computation (VUBQC) circumvents these shortcomings by utilising cryptographic techniques, treating deviations and noise as potential security threats \cite{Fitzsimons_2017, Broadbent_2009, leichtle2021verifyingbqpcomputationsnoisy, Gheorghiu_2018}. By physically separating the verifier's trusted small quantum device (e.g. random single qubit generator) from the large QPU to be tested, the full computation can be obfuscated without revealing more than the size of the computation to the QPU. However, such schemes require the transfer of quantum information between the verifier and QPU. So far all proof of concept demonstrations of such schemes have used academic labs with networking infrastructure \cite{Drmota_2024, Barz_2013}. 

Obfuscating the computation enforces that any malicious behaviour of the Server (any general noise of the QPU) is independent of the description of the computation. This in turn guarantees that the behaviour is the same between interleaved test rounds, which can be verified, and computation rounds performing the desired computation. To our knowledge, this is the only scalable, provable, robust way of verifying an arbitrary quantum computation against the most general noise model. However, the transfer of quantum information adds significant noise and technological overhead, making this class of protocols an extremely challenging test for emerging QPUs to pass. Until interconnected QPUs become commonplace alternative solutions to enforce the requirements of VUBQC on-chip should be found. 

We propose and demonstrate such a scheme, called on-chip verification, replacing a fully malicious server with an honest but potentially noisy device; matching closely the scenario facing most QPU providers at present. We assume all the classical control can be trusted, and that all the verifier's instructions are followed. At the cost of blindness, we remove the need to physically separate verifier and prover, but maintain the independence of the Server's actions from the type of round, compute or test, that it is implementing.  We make no assumption on the noise other than that it is a CPTP map independent of the secret parameters used to prepare the input single qubit states. We provide a methodology for measuring this secret-dependency which, due to the robustness property of the protocol, will introduce a bound on the confidence of the accepted computation.

While other proof of principle classical client verification schemes have been demonstrated \cite{Lewis_2024}, a significant qubit overhead would be required to meet the computational assumptions of these schemes. Protocols employing an information-theoretically secure cryptographic approach have also been demonstrated using photons to transfer data \cite{Drmota_2024}. In that work, noise levels were identified as too high to scale the computation beyond a few qubits. Accreditation \cite{ferracin2019accrediting} is another gate-based approach to verification which attempts to replace cryptographic techniques with assumptions on the QPU's noise. However, the noise assumptions required to apply accreditation are strong compared to our minimal requirement on the noise model. 

We demonstrate that our desired noise conditions are closely met by the Quantinuum H1-1 Quantum Charge-Coupled Device (QCCD) trapped ion device, made possible by the low cross-talk isolated gate zones. Further, the availability of mid-circuit measurements and resets on the H1-1 device, and the capacity to perform complicated classical expressions within the coherence time of the qubits, are all necessary to implement our protocol. These features, in addition to the all-to-all connectivity of the qubits, and low measurement and gate error rates, allow us to demonstrate the largest successful verified measurement-based quantum computation pattern to date, constructing a 52-vertex MBQC graph state.

This shows that most of the building blocks for VUBQC are well aligned with the required components of Fault-Tolerant Quantum Computing. However, per-shot randomness is commonly unavailable in any software stack. We put this forward as a requirement for the quantum computing stack, but in the mean time present and justify a route to recreate this feature.

In \cref{sec:background}, we discuss existing verification schemes and present-day QPU technology. In \cref{sec:qccd verification protocol}, we introduce our approach to the verification of quantum computations on near-term QPUs, making assumptions on secret independence which we justify in \cref{sec:assesments}. We present results and details of several implementations of our approach in \cref{sec:verification experiment} before concluding in \cref{sec:conclusion}.

\section{Background}
\label{sec:background}

In \cref{sec:mbqc} we introduce Measurement-Based Quantum Computing, which is the model of computation used by our on-chip verification protocol. In \cref{sec:verification}, we introduce verified quantum computing, and in particular \cite{leichtle2021verifyingbqpcomputationsnoisy}, to support the description of our protocol in \cref{sec:qccd verification protocol}. In \cref{sec:qccd}, we describe the QCCD architecture and the H1-1 device which we will use in our experiments. The characteristics of the QCCD architecture will influence the design of our protocol in \cref{sec:qccd verification protocol} and our experiments in \cref{sec:verification experiment}.

\subsection{Measurement-Based Quantum Computation}
\label{sec:mbqc}

Measurement-Based Quantum Computation (MBQC)~\cite{raussendorf2001one} is an alternative framework to the conventional gate-based model for quantum computation. MBQC executes computations by adaptively measuring a highly entangled graph state. An MBQC algorithm is defined by a set $\{(G, I, O), \phi, f\}$. $(G, I, O)$ represents an \emph{open graph}, consisting of a connected graph $G=(V,E)$ with vertices $V$ and edges $E$, and input and output sets $I$ and $O$. The vertices $V$ represent qubits, and edges $E$ correspond to entangling operations between the connected qubits. $\phi$ is the set of measurement angles for each measured vertex. The function $f: O^c \rightarrow I^c$, known as the \emph{flow}~\cite{danos2006determinism}, is an injective mapping from the non-output set $O^c$ to the non-input set $I^c$. This flow defines the corrections required to achieve deterministic computation on the graph state.

Single-qubit state preparations and measurements are typically performed on the same fixed plane of the Bloch sphere. In this work, we use the YZ-plane. State preparations on the YZ plane are defined:
\begin{equation}\label{eq:state_preparation}
\ket*{\theta}_{\mathrm{YZ}}\equiv \ket{\theta} =\cos\frac{\theta}{2}\ket{0}
-i\sin{\frac{\theta}{2}}\ket{1}.
\end{equation}
A measurement by angle $\theta$ corresponds to applying the measurement projectors $\{\ketbra{\theta},\ketbra{{\theta+\pi}}\}$.

The process of preparing graph states varies depending on the selected plane. Each vertex corresponds to a state prepared in its respective plane as defined in \cref{eq:state_preparation}. In the YZ-plane, an edge corresponds to applying $\mathrm{XX}_{ij}=H_iH_j\mathrm{CZ}_{ij}H_i H_j$, where $H$ indicates the Hadamard gate and $\mathrm{CZ}$ is the controlled-$Z$ gate. Thus, the graph state is stabilised by stabilisers $\{Z_v\bigotimes_{j\in N_G(v)} X_j\}_{v\in V}$, where $N_G(v)$ denote the set of neighbouring vertices of $v$.

\subsection{Verification of Quantum Computation}
\label{sec:verification}

With the emergence of delegated quantum computing as a service, verifying the correctness of the results of a quantum computation becomes a task of crucial importance. Verification protocols give the Client, also called \emph{verifier}, the power to recognise corrupted results returned by a noisy or malicious quantum server, also called \emph{prover}. Recent advances \cite{leichtle2021verifyingbqpcomputationsnoisy,Kapourniotis_2024,kapourniotis2023asymmetricquantumsecuremultiparty,kashefi2024verification} in the field have drastically improved the efficiency over earlier techniques and protocols \cite{Fitzsimons_2017,Kashefi_2017,Gheorghiu_2018}, leading to the first experimental implementations of verified quantum computations \cite{Drmota_2024,polacchi2024experimentalverifiablemulticlientblind}.

\paragraph{Verification with Trapification.}
The most resource-efficient verification protocol to date~\cite{leichtle2021verifyingbqpcomputationsnoisy} employs trapification techniques for embedded error detection in conjunction with classical repetition codes. This combination allows the secure delegation of arbitrary BQP computations to an untrusted server with an efficient number of repetitions as the only overhead. Additionally, the protocol is robust even to a constant global noise. These properties make verification a promising possibility on near-term devices.

The verification protocol proposed by \cite{leichtle2021verifyingbqpcomputationsnoisy} works by interleaving computation rounds, implementing randomly compiled versions
of the target computation, with test rounds, which detect any error that has the potential to harm the computation results. Indistinguishability between the different rounds guarantees that the assessment of the noise made by the test rounds is representative of the errors occurring in computation rounds.

\paragraph{Relation to benchmarking.}
The verification protocol demonstrated in this work can be reinterpreted as a benchmark of quantum devices. In general, blind quantum verification protocols that achieve composable security must have the property of \emph{independent verifiability}~\cite{Dunjko_2014}, which is to say that the Client's decision to accept or reject at the end of the protocol should be independent of the delegated computation\footnote{The decision whether the Client accepts is still allowed to depend on the resources used for the delegated computation, \emph{i.e.}, the graph that describes the resource state as well as the order in which the qubits are processed.}. After the successful execution of a protocol with such security, the Client will have not only learnt the correct result of the target computation evaluated on their input, but also that the protocol would have equally accepted (with the same probability) on the input of any other target computation consuming the same resources. In this way, the Client obtains side information about the computational power of the Server's quantum device.

This observation led to the design of verification-inspired benchmarking schemes~\cite{frank2024heuristicfreeverificationinspiredquantumbenchmarking}, which reinterpret verification routines as cryptographic quantum benchmarks. One significant advantage of this novel approach to benchmarking is the absence of heuristics, allowing it to make strong predictive statements about the computational power of tested quantum computers. Benchmarking schemes which follow this design paradigm are inherently scalable, a crucial feature that many previously proposed benchmarking methods fail to attain.

\paragraph{The on-chip setting.}
All known information-theoretically secure single-prover quantum verification schemes require the verifier to operate coherently on some low-dimensional quantum states, often by preparing or measuring single-qubit states in variable bases. This requirement comes with the implicit assumption that the verifier has access to a device that can perform these operations and that the device is trusted, which implies noise-free and perfect operations in cryptographic terms. Additionally, these protocols require quantum communication between Client and Server, which renders them infeasible on hardware architectures which do not support quantum interconnections, or in settings that lack the necessary infrastructure to exchange coherently quantum information between the two parties. Therefore, there is a need for protocols with entirely classical clients, without introducing prohibitive overheads from the evaluation of post-quantum cryptographic primitives in superposition \cite{mahadev2023classicalverificationquantumcomputations,Cojocaru_2019}.

In Section~\ref{sec:qccd verification protocol}, we present a way of reformulating verification schemes from the usual prepare-and-send setting between Client and Server to a setting in which all quantum operations are performed on a single chip on the Server's side, which we will refer to as the \emph{on-chip setting}. Rather than relying on computational assumptions and introducing new hardware overheads, we pay for removing quantum communication with assumptions on the behaviour of the Server's quantum device, which can be seen as a form of assumption on the noise structure. In particular, the client-server separation in the usual cryptographic setting translates to assumptions on the secret dependency of the noise during single-qubit preparation on the Server's quantum machine and the faithfulness of the Server's classical controls in the on-chip setting. For more details on the concept of secret dependency, see~\Cref{sec:secret_dependency}.

\subsection{Trapped-Ion QCCD}
\label{sec:qccd}

We demonstrate that a QCCD ion trap has the capabilities to perform MBQC with minimal quantum resources. It also has the potential to meet the requirements of our security assumptions, namely, isolation during gate operations and circuit branching\footnote{We refer to ``circuit branching'' as decisions in the circuit execution which depend on classical variables that are generated on the fly, e.g., previous measurement outcomes.} per measurement shot. Therefore, this system is one of the best choices for our on-chip verification protocol.

A typical QCCD trap consists of segmented electrodes with high-precision voltage control, enabling ion trapping and high-fidelity physical transport. Some segments, or \emph{zones}, are addressed by external lasers to implement qubit operations, e.g., qubit initialisation, gates, and measurement. Some segments may also have additional electrodes to allow physical swap of ions to rearrange and pair different qubits. These key features enable all-to-all connectivity, parallel operations \cite{moses2023race}, high-fidelity two-qubit gates, and isolated operations, enabling mid-circuit measurement and reset, while exhibiting low crosstalk \cite{gaebler2016high, ballance2016high, Pino_2021, gaebler2021suppression}.

In this paper, we perform our experiments on the Quantinuum trapped ion H1-1 device \cite{Pino_2021}  and its noisy classical emulator H1-1E \cite{QuantinuumH1Emulator2024, pecos, crathesis}. The H1-1 trap features five interaction zones \cite{quantinuum_h_series}, where the quantum operations -- including initialisation, measurement, quantum gates, and cooling -- are performed using lasers. The trap is integrated with FPGAs, enabling classically conditioned circuit branching per measurement shot. 

The H1-1 device contains 20 qubits, defined within the atomic hyperfine states of ${}^{171}\mbox{Yb}^{+}$ \cite{olmschenk2007manipulation,mai2024high}. Sympathetic cooling \cite{barrett2003sympathetic} is actively performed throughout the computation, by pairing the ion with a ${}^{138}\mbox{Ba}^{+}$ coolant. In particular, doppler cooling~\cite{eschner2003laser} and resolved sideband cooling~\cite{monroe1995resolved} are performed. Such a qubit configuration facilitates extremely long coherence times, e.g., exceeding 10 minutes \cite{wang2017single}.

A quantum circuit is compiled into the H1-1 set of native gate operations, including pairing, isolation, and physical transport. The compilation process is optimised by the minimal number of physical transports. The native single-qubit gate is achieved via stimulated Raman transitions, and is defined as
\[
    U(\theta,\phi) = e^{-i \frac{\theta}{2} (\cos\phi X+\sin\phi Y )}.
\]
The native two-qubit gate is implemented using the M{\o}lmer-S{\o}rensen interaction \cite{sorensen2000entanglement} in the phase sensitive configuration \cite{lee2005phase}, defined as
\[
    \mathrm{ZZ}(\theta) =  e^{-i \frac{\theta}{2}(Z\otimes Z)}.
\]
Additionally, the phase gate 
\[
    \mathrm{Rz}(\theta) = e^{-i \frac{\theta}{2} Z},
\]
is performed virtually \cite{mckay2017efficient}, which is to say it is tracked in software.

\section{On-Chip Verification and Benchmarking Protocols}
\label{sec:qccd verification protocol}

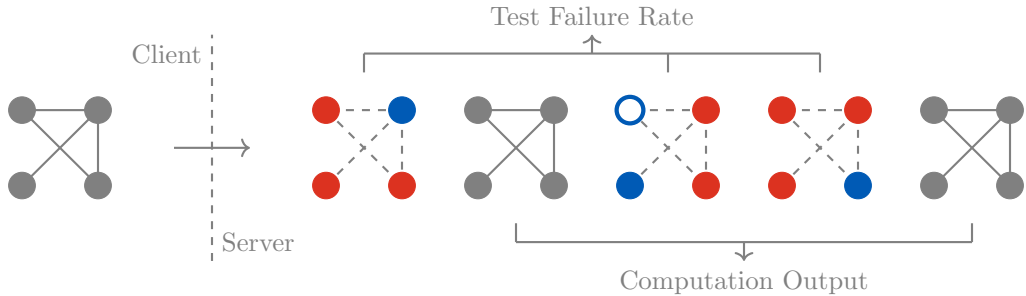
\begin{figure*}
    \centering
    \begin{tikzpicture}

        \definecolor{trap}{HTML}{005AB5}
        \definecolor{dummy}{HTML}{DC3220}

        \draw[thick, gray] (0,10) -- (1,11);
        \draw[thick, gray] (1,10) -- (1,11);
        \draw[thick, gray] (1,11) -- (0,11);
        \draw[thick, gray] (0,11) -- (1,10);
    
        \filldraw [gray] (0,10) circle (5pt);
        \filldraw [gray] (1,10) circle (5pt);
        \filldraw [gray] (0,11) circle (5pt);
        \filldraw [gray] (1,11) circle (5pt);


        \draw[thick, gray, dashed] (2.5,9) node[anchor=south west]{Server}  -- (2.5,12) node[anchor=north east]{Client};

        \draw[thick, gray, ->] (2,10.5) -- (3,10.5);


        \draw[thick, gray, dashed] (4,10) -- (5,11);
        \draw[thick, gray, dashed] (5,10) -- (5,11);
        \draw[thick, gray, dashed] (5,11) -- (4,11);
        \draw[thick, gray, dashed] (4,11) -- (5,10);
    
        \filldraw [dummy] (4,10) circle (5pt);
        \filldraw [dummy] (5,10) circle (5pt);
        \filldraw [dummy] (4,11) circle (5pt);
        \filldraw [trap] (5,11) circle (5pt);


        \draw[thick, gray] (6,10) -- (7,11);
        \draw[thick, gray] (7,10) -- (7,11);
        \draw[thick, gray] (7,11) -- (6,11);
        \draw[thick, gray] (6,11) -- (7,10);
    
        \filldraw [gray] (6,10) circle (5pt);
        \filldraw [gray] (7,10) circle (5pt);
        \filldraw [gray] (6,11) circle (5pt);
        \filldraw [gray] (7,11) circle (5pt);


        \draw[thick, gray, dashed] (8,10) -- (9,11);
        \draw[thick, gray, dashed] (9,10) -- (9,11);
        \draw[thick, gray, dashed] (9,11) -- (8,11);
        \draw[thick, gray, dashed] (8,11) -- (9,10);
    
        \filldraw [trap] (8,10) circle (5pt);
        \filldraw [dummy] (9,10) circle (5pt);
        \filldraw [color=trap, fill=white, ultra thick] (8,11) circle (5pt);
        \filldraw [dummy] (9,11) circle (5pt);
        

        \draw[thick, gray, dashed] (10,10) -- (11,11);
        \draw[thick, gray, dashed] (11,10) -- (11,11);
        \draw[thick, gray, dashed] (11,11) -- (10,11);
        \draw[thick, gray, dashed] (10,11) -- (11,10);
    
        \filldraw [dummy] (10,10) circle (5pt);
        \filldraw [trap] (11,10) circle (5pt);
        \filldraw [dummy] (10,11) circle (5pt);
        \filldraw [dummy] (11,11) circle (5pt);


        \draw[thick, gray] (12,10) -- (13,11);
        \draw[thick, gray] (13,10) -- (13,11);
        \draw[thick, gray] (13,11) -- (12,11);
        \draw[thick, gray] (12,11) -- (13,10);
    
        \filldraw [gray] (12,10) circle (5pt);
        \filldraw [gray] (13,10) circle (5pt);
        \filldraw [gray] (12,11) circle (5pt);
        \filldraw [gray] (13,11) circle (5pt);


        \draw[thick, gray] (4.5, 11.75) -- (10.5, 11.75);
        \draw[thick, gray] (4.5, 11.75) -- (4.5, 11.5);
        \draw[thick, gray] (8.5, 11.75) -- (8.5, 11.5);
        \draw[thick, gray] (10.5, 11.75) -- (10.5, 11.5);
        \draw[thick, gray, ->] (7.5, 11.75) -- (7.5, 12) node [anchor=south] {Test Failure Rate} ;

        \draw[thick, gray] (6.5, 9.25) -- (12.5, 9.25);
        \draw[thick, gray] (6.5, 9.25) -- (6.5, 9.5);
        \draw[thick, gray] (12.5, 9.25) -- (12.5, 9.5);
        \draw[thick, gray, ->] (9.5, 9.25) -- (9.5, 9) node [anchor=north] {Computation Output} ;
        
    \end{tikzpicture}
    \caption{\textbf{Outline of On-Chip Verified Quantum Computing, as described in Protocol \ref{prot:on-chip-VBQC}.} The Client generates a description of an MBQC computation and communicates it to the Server. The Server randomly alternates between performing computation rounds and test rounds. Computation rounds, seen in gray, are randomised versions of the Client's computation, with the same ideal output. By randomly compiling the computation rounds the actions of the Server are indistinguishable between rounds. The output from the computation rounds is then ideally the original computation output. During test rounds a random colour from a colouring of the graph is selected. Coloured vertices are assigned as test vertices, seen in blue, and are disentangled using dummy vertices, seen in red. The measurement outcomes of test vertices are deterministic in the ideal case. Test rounds where any measurement is found to be incorrect, indicated with unfilled vertices, is said to have failed. The output of the test rounds is the test failure rate, which, as the operations performed by the Server are indistinguishable between rounds, gives a measure of the confidence in the computation output. See \cref{sec:compilation} and \cref{fig:classical registers} for details on how the corresponding qubits are generated on H1-1.}
    \label{fig:protocol outline}
\end{figure*}

Our on-chip verification protocol, described in Protocol \ref{prot:on-chip-VBQC} and outlined in \cref{fig:protocol outline}, adapts the noise-robust verification protocol from \cite[Protocol~1]{leichtle2021verifyingbqpcomputationsnoisy}, moving the preparation of the randomized single-qubit states from the Client's to the Server's device. The prepared single-qubit states at the beginning of the protocol are chosen from the set $\{ \ket{\theta} \,|\, \theta \in \Theta \} \cup \{ \ket{+}, \ket{-} \}$, where $\Theta = \{ \frac{j\pi}{4} \,|\, j=0,\dots,7 \}$ and $\ket{\theta} = \mathrm{Rx} (\theta)\ket{0}$ (\cref{eq:state_preparation}). While previously proposed verification protocols use states in the XY-plane, Protocol \ref{prot:on-chip-VBQC} is formulated using rotated states in the YZ-plane, as this is the appropriate choice for the H1-1 machine.

\begin{figure*}[t]  	
\begin{minipage}{\textwidth}
\begin{algorithm}[H]
\caption{\raggedright On-chip verification of BQP computations}
\label{prot:on-chip-VBQC}
\begin{algorithmic}[0]
\STATE \textbf{Parameters:} Number of computation rounds $d$, number of test rounds $t$, total number of rounds $n=d+t$, flow $f$ on graph $G=(V,E)$, and a $k$-coloring $K$ of $G$.
\STATE \textbf{Client's Inputs:} Angles $\qty{\phi_v}_{v \in V}$, classical input to the computation $x \in \bin^{\#I}$ (where $\#I$ is the size of the set $I$ of input vertices).
\STATE \textbf{Protocol:}
\begin{enumerate}
\item The Client chooses uniformly at random a partition $(C, T)$ of $[n]$ ($C \cap T = \emptyset$) with $\#C = d$, the sets of indices of the computation and test rounds respectively.

\item[2.] For $j \in [n]$, the Client and the Server perform the following sub-protocol (the Client may send message $\Redo_j$ to the Server before step 2.c while the Server may send it to the Client at any time, both parties then restart round $j$ with fresh randomness):
\begin{enumerate}

\item[(a)] If $j \in T$ (test), the Client chooses uniformly at random a colour $\mathsf{V}_j \sample \qty{V_k}_{k \in [K]}$ (this is the set of traps for this test round).
\item[(b)] The Client instructs the Server to prepare $\#V$ qubits. If $j \in T$ and the destination qubit $v \notin \mathsf{V}_j$ is a non-trap qubit (therefore a dummy), then the Client chooses uniformly at random $d_v \sample \bin$ and instructs the preparation of the state $Z^{d_v}\ket{+}$. Otherwise, the Client chooses at random $\theta_v \sample \Theta$ and instructs the preparation of the state $\ket{{\theta_v}}$.
\item[(c)] The Server performs a $\mathsf{X}\mathsf{X}$-gate between all its qubits corresponding to an edge in the set $E$.
\item[(d)] For $v \in V$, the Client sends a measurement angle $\delta_v$, the Server measures the appropriate corresponding qubit in the $\delta_v$-basis of the $\mathsf{Y}\mathsf{Z}$-plane, returning outcome $b_v$ to the Client. The angle $\delta_v$ is defined as follows:
\begin{itemize}
\item If $j \in C$ (computation), it is the same as in UBQC, computed using the flow and the computation angles $\qty{\phi_v}_{v \in V}$. For $v \in I$ (input qubit) the Client uses $\tilde{\theta}_v = \theta_v + x_v\pi$ in the computation of $\delta_v$.
\item If $j \in T$ (test): if $v \notin \mathsf{V}_j$ (dummy qubit), the Client chooses it uniformly at random from $\Theta$; if $v \in \mathsf{V}_j$ (trap qubit), it chooses uniformly at random $r_v \sample \bin$ and sets $\delta_v = \theta_v + r_v\pi$.
\end{itemize}
\end{enumerate}

\item[3.] For all $j \in T$ (test round) and $v \in \mathsf{V}_j$ (traps), the Client verifies that $b_v = r_v \oplus d_v$, where $d_v = \bigoplus_{i \in N_{G}(v)} d_i$ is the sum over the values of neighbouring dummies of qubit $v$. Let $c_{\mathit{fail}}$ be the number of failed test rounds (where at least one trap qubit does not satisfy the relation above), if $c_{\mathit{fail}} \geq w$ then the Client aborts by sending message $\Abort$ to the Server.

\item[4.] Otherwise, let $y_j$ for $j \in C$ be the classical output of computation round $j$ (after corrections from measurement results). The Client checks whether there exists some output value $y$ such that $\# \left\{ y_j \, | \, j \in C,\, y_j = y \right\} > \frac{d}{2}$. If such a value $y$ exists (this is then the majority output), it sets it as its output and sends message $\Ok$ to the Server. Otherwise it sends message $\Abort$ to the Server.
\end{enumerate}
\end{algorithmic}
\end{algorithm}
\end{minipage}
\end{figure*}

\cref{thm:security-on-chip-verif} captures the security of Protocol~\ref{prot:on-chip-VBQC} and formalises the assumptions necessary for on-chip verifiability. Note, that Theorem~\ref{thm:security-on-chip-verif} makes no claim about the blindness of Protocol~\ref{prot:on-chip-VBQC}. The notion of blindness is not applicable in our setting where the Server has full access to all previously secret information and parameters of the protocol. Security of the protocol therefore solely refers to the verifiability of the results, limited to assumptions on the behaviour of the Server's classical and quantum devices.

\begin{theorem}[Security of on-chip verification protocol]\label{thm:security-on-chip-verif}
    For $n = d+t$ such that $d/n$ and $t/n$ are fixed in $(0,1)$ and $w$ such that $w/t$ is fixed in $(0, \frac{1}{k}\cdot \frac{2p-1}{2p-2})$, where $p$ is the inherent error probability of the BQP computation, Protocol~\ref{prot:on-chip-VBQC} with $d$ computation rounds, $t$ test rounds, and a maximum number of tolerated failed test rounds of $w$ achieves $\varepsilon$-verifiability with $\varepsilon$ exponentially small in $n$ under the following assumptions:
    \begin{enumerate}
        \item All classical computations and controls on the Server's side are operated honestly, \textit{i.e.}, the Server's quantum device receives the correct gate instructions.
        \item The noise during the initial single-qubit preparation is secret-independent, \textit{i.e.}, the actual preparation of the states can be simulated by applying a noise channel to perfectly prepared states which is independent of their classical description.
    \end{enumerate}
\end{theorem}

\cref{prot:on-chip-VBQC} differs from \cite[Protocol~1]{leichtle2021verifyingbqpcomputationsnoisy} only in the initial preparation of the single-qubit states, which is performed on the Server's quantum device rather than on the Client's. In the following, we show that the assumed secret-independence of the initial preparation noise in \cref{prot:on-chip-VBQC} means that this noise could just as well have been simulated by the Server in \cite[Protocol~1]{leichtle2021verifyingbqpcomputationsnoisy}. This reduces the security of \cref{prot:on-chip-VBQC} to the security of \cite[Protocol~1]{leichtle2021verifyingbqpcomputationsnoisy}.

\begin{proof}[Proof sketch.]

    Assumption~2 in Theorem~\ref{thm:security-on-chip-verif} guarantees that the noise during these preparations can be simulated by the Server using a secret-independent noise channel after perfect preparations. In this way, errors due to the noise during the preparation phase are accounted for in the traditional security analysis of the verification protocol which does not restrict the Server's deviations from the protocol instructions as long as they are independent of the Client's secrets $d_v, \theta_v, r_v$.
    In this way, \cref{prot:on-chip-VBQC} is equivalent to a protocol in which the single-qubit preparation is performed on a trusted and perfect device which is separated from the Server and directly instructed by the Client.

    Assumption~1 in Theorem~\ref{thm:security-on-chip-verif} guarantees that no other operations on the Server's device depend on the Client's secrets. Consequently, the Server in this hybrid protocol becomes entirely secret-independent which maps any potential attack on Protocol~\ref{prot:on-chip-VBQC} satisfying above assumptions back to the original verification protocol from \cite{leichtle2021verifyingbqpcomputationsnoisy} in a setting with physical separation between the trusted Client and the untrusted Server.

    The application of~\cite[Theorem~1]{leichtle2021verifyingbqpcomputationsnoisy} then implies the claim.
\end{proof}

As one example application of Theorem~\ref{thm:security-on-chip-verif}, in the case of deterministic computations with inherent error probability $p=0$ and $k=2$ different types of test rounds, the acceptable threshold for the choice of $w/t$ equals $25\%$. Hence, also the maximum test failure rate due to noise that the protocol can be made robust to is $25\%$. In the experiments of \cref{sec:verification experiment} $k=2$ is the relevant parameter, and so we will use $25\%$ test failure rate as a benchmark.

Following the benchmarking paradigm of~\cite{frank2024heuristicfreeverificationinspiredquantumbenchmarking}, the proposed verification scheme can straightforwardly be turned into a benchmarking scheme. The main modification is the omission of computation rounds, and a slightly adapted classical postprocessing of the quantum measurements. As such, the resulting benchmarking scheme is mainly a reinterpretation of the Server's responses during the verification protocol. This allows us in the following to implement and perform both protocols at once -- verifying a specific computation while benchmarking the quantum device at hand with respect to fixed resources.

Passing this benchmark certifies that the Server's quantum device is able to correctly evaluate any computation from the class of computations that consume at most a certain amount of resources, \textit{i.e.}, a restricted number of qubits, and a fixed graph state as entangled resource state.

\section{Assessing Protocol
Requirements}
\label{sec:assesments}

In \cref{sec:qccd verification protocol} we described our protocol for verified quantum computation on near-term QCCD architectures. Two key security requirements of our protocol are that any noise during the initial single-qubit state preparation is independent of the secrets, and the availability of high-quality random numbers.

The secret independence of noise during the initial state preparations naturally holds in the traditional client-server setups. This is due to their spatial separation and the explicit assumption that the Client is operating an honest and perfect device. However, since our approach eliminates this separation, we must directly evaluate such noise secret dependency, as detailed in \cref{sec:secret_dependency}.

Fresh random numbers are required for each measurement shot, with circuit branching determined by these values. We generate these random numbers on the fly by measuring Hadamard basis states in the computational bases. We assess the quality of our Random Number Generator (RNG) in \cref{sec:random_numbers}.

\subsection{Secret-Independency}
\label{sec:secret_dependency}

Recall that noise during the preparation process of \cref{prot:on-chip-VBQC} which do not depend on the protocol secrets remains harmless, as it could have happened equally during the later adversarial stage of the protocol. Importantly, this notion of \emph{secret-independency} is robust. Specifically, if the noise during the state preparation is close in trace distance to a secret-independent noise channel, it can degrade the security guarantees of the protocol only by an amount proportional to this distance. This follows due to the indistinguishability of the realistically prepared states from a family of states suffering at most from harmless, secret-independent noise.

In this work, we implement MBQC on the YZ-plane with state preparations as described in \cref{eq:state_preparation}. We selected the YZ-plane over the standard XY-plane because, in the XY-plane, the randomisations (state preparation and measurement basis transformation) operations primarily consist of $Z$-rotations, which are virtual operations managed by software on H1-1. Such randomisation in the YZ-plane can be done with one rotation around the $X$-axis.

The secret dependency is quantified by
\begin{equation}\label{eq:secret_dependent}
    \min_{\mathcal E} \frac{1}{\abs{\Theta}}\sum_{\theta\in\Theta}{\|\mathcal E (\ketbra{{\theta}})-\rho_{\theta}\|_1 \coloneqq \min_{\mathcal E}\Delta(\mathcal E)},
\end{equation}
where $\theta$ is the secret, $\ket{{\theta}}$ is the noiseless state preparation, $\rho_{\theta}$ is the prepared state in the quantum computer, and $\|.\|_1$ is Schatten-1 norm (also known as the trace norm). Thus, the noise is independent of the secret if there exists a CPTP map $\mathcal E$ such that $\Delta(\mathcal E) = 0$. $\Delta(\mathcal E)$ then quantifies distinguishability between resources in the ideal and the real world. 

We perform single-qubit tomography on eight quantum states $\{\rho_\theta\}$ where $\theta\in\{j\pi/4\}_{j\in[0,7]}$. We perform 3000 shots of measurement on each Pauli-measurement, for each $\theta$. We then perform bootstrapping to approximate the effect of sampling error by resampling 1,000 sets of measurements, each with a chunk size of 1,000 measurements. \cref{tab:infidelity} presents the infidelities associated with preparing $\rho_\theta$. 

\begin{table}[tphb]
\resizebox{0.9\columnwidth}{!}{
\begin{tabular}{|c|c|c|c|c|c|c|c|c|}
\hline
$\theta$ &
0 & $\frac{\pi}{4}$ &  $\frac{\pi}{2}$ & $\frac{3\pi}{4}$ & $\pi$ & $\frac{5\pi}{4}$ & $\frac{3\pi}{2}$ & $\frac{7\pi}{4}$  \\ \hline
$\mu$&
5.3&4.4&6.0&4.5&8.2&7.3&4.8&5.8\\ \hline
var&
2.9&1.8&3.8&2.3&8.2&4.8&2.2&3.4\\ \hline
\end{tabular}
}~\hspace{-1em}
\begin{tabular}{l}
\\ \small$\times10^{-4}$ \\
\small$\times10^{-7}$ \\
\end{tabular}
    \caption{Average infidelity $(\mu)$ and variance (var) of state preparations $\ket{\theta}$ on the YZ-plane through bootsrapping. The average infidelity overall is $5.8\times 10^{-4}$ or fidelity 0.9994.}
    \label{tab:infidelity}
\end{table}

We employ the Frobenius norm $\|.\|_F$ instead of the trace norm for the optimisation in \cref{eq:secret_dependent}. It is chosen because \cref{eq:secret_dependent} can be directly formulated as a convex optimisation problem under the Frobenius norm, as the norm satisfies the triangle inequality: $\|A+B\|_F\leq\|A\|_F+\|B\|_F$. The Frobenius norm is more practical for optimisation purposes, as it is computed directly from the matrix entries, unlike the trace norm, which requires matrix decompositions.
Finally, we use \texttt{ConvexOptimization} function of \texttt{Mathematica}~\cite{wolfram2024convexoptimization}, utilising MOSEK~\cite{mosek} solver. The details of this optimisation are elaborated on in \cref{sec:optimisation}.

Despite the high fidelity of state preparation ($\approx$ 0.9994), our analysis reveals that the noise in the state preparation is not completely secret independent. Using the optimisation under Frobenius norm, we obtain a map $\tilde{\mathcal E}$ with objective value $\varepsilon_F=0.015$, where 
\begin{equation}
\tilde{\mathcal E}_P=
\begin{pmatrix}
1 & 0 & 0 & 0 \\
-0.002 & 0.998& 0.006& 0.003 \\
-0.001 & -0.007& 0.998& 0.003 \\
0 & -0.004& -0.002 & 0.999
\end{pmatrix}
\end{equation}
in the Pauli transfer matrix representation. The diagonal elements of $\mathcal{E}_P$ suggest characteristics similar to dephasing noise, while the block diagonal structure indicates the presence of rotation errors. We elaborate on the details of our optimisation and tomography results in \cref{sec:optimisation}.

Using the same map $\tilde{\mathcal E}$, we obtain a trace norm value of $\varepsilon_1=0.011$ (\cref{thm:security-on-chip-verif}). This shows that the security guarantees of our protocol are not degraded more by the secret-dependency of the initial single-qubit state preparation than $\varepsilon_1$ multiplied by the number of qubit preparations. This remains a potentially loose bound; we leave the question of tighter bounds for future research.

\subsection{Per-Shot Randomness}
\label{sec:random_numbers}

\cref{prot:on-chip-VBQC} requires multiple instances of randomness for each measurement shot and each vertex in the graph state. However, these requirements are commonly not found in the software stack of a quantum computer system. To overcome this problem, we embed our RNG within the circuit. This configuration is detailed in \cref{sec:compilation} and illustrated in \cref{fig:classical registers}.   

Our random numbers are generated by preparing $\ket{+}$ states and measuring them in the computational basis. To assess the quality of our RNGs, we generate random bits and subject them to a standard test, specifically the Federal Information Processing Standard (FIPS) 140-2~\cite{nist140-2}. FIPS 140-2 is developed by the National Institute of Standards and Technology (NIST), and defines security requirements for cryptographic modules used by both government and industry to protect sensitive information.
The tests provide explicit statistical bounds that the computed values must meet. 

A single bit stream of 20,000 random bits was subjected to four tests: the Monobit test, the Poker test, the Run test, and the Long Run test. We utilise an open-source software, \texttt{rng-tools}~\cite{rng_tools},  to execute these tests, all of which our random bits have successfully passed.
Detailed descriptions of these statistical tests can be found in references~\cite{nist140-2,fdk_hm-rae103}. The sequence of bits used is available as detailed in `Data Availability'.

\section{On-Chip Verification Experiments}
\label{sec:verification experiment}

\cref{sec:compilation} describes how on-chip verified computation is implemented on the H1-1 device. We describe the results of two verified computations implemented in this way: in \cref{sec:two qubit grover}, a two qubit Grover's search algorithm; and in \cref{sec:cnot grid}, a larger measurement pattern implementing a grid of CNOTs.

In those results sections we will study two quantities, as outlined in \cref{fig:protocol outline}:
\begin{description}
    \item[Test failure rate:] The fraction of the test rounds where at least one trap fails.
    \item[Incorrect output rate:] The fraction of the computation rounds where the incorrect output is produced.
\end{description}
Note that it is possible to calculate the incorrect outcome rate in these cases as the ideal correct outcome is known and deterministic; this is for demonstration purposes only and is not required in the protocol. The 2-qubit Grover's search pattern is small enough that the solution is known, but uses components which when scaled would not be efficiently classically simulatable. The CNOT grid implements a classical circuit so can be used as a scalable benchmark of verified quantum computation. However the CNOT grid experiment measurement pattern uses a number of qubits far larger than any other verified MBQC computation implemented to date.

\subsection{H1-1 QCCD Compilation}
\label{sec:compilation}

The verification technique described in \cref{sec:verification} utilises random state preparation. The random state prepared depends firstly on the type of round, being either a computation or a test round, with the round itself being selected at random. In the case of the test rounds the state prepared depends secondly on a randomly selected assignment of the vertices as being either a trap or dummy. This is to say a randomly selected colour from a colouring of the graph. Finally, in the case of both the test round and the computation round, the state prepared depends on a random one time pad of the measurement angles and the measurement outcomes.

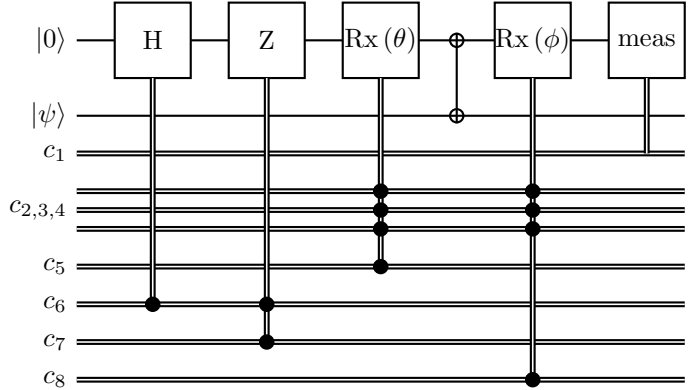
\begin{figure*}
    \centering
    \begin{tikzpicture}[thick, scale=0.5]

        \draw (-1,10) node[anchor=east]{$\left| 0 \right\rangle$} -- (13,10);

        \draw (-1,8) node[anchor=east]{$\left| \psi \right\rangle$} -- (15,8);

        \draw[double] (-1,7) node[anchor=east]{$c_1$} -- (15,7);

        \draw[double] (-1,6) -- (15,6);
        \draw[double] (-1,5.5) node[anchor=east]{$c_{2,3,4}$} -- (15,5.5);
        \draw[double] (-1,5) -- (15,5);

        \draw[double] (-1,4) node[anchor=east]{$c_5$} -- (15,4);

        \draw[double] (-1,3) node[anchor=east]{$c_6$} -- (15,3);

        \draw[double] (-1,2) node[anchor=east]{$c_7$} -- (15,2);

        \draw[double] (-1,1) node[anchor=east]{$c_8$} -- (15,1);

        \filldraw[fill=white] (0,9) rectangle (2,11) node[pos=.5]{$\mathrm{H}$};
        \draw[double] (1,9) -- (1,3);
        \filldraw[black] (1,3) circle (5pt);

        \filldraw[fill=white] (3,9) rectangle (5,11) node[pos=.5]{$\mathrm{Z}$};
        \draw[double] (4,9) -- (4,2);
        \filldraw[black] (4,3) circle (5pt);
        \filldraw[black] (4,2) circle (5pt);

        \filldraw[fill=white] (6,9) rectangle (8,11) node[pos=.5]{$\mathrm{Rx}\left(\theta\right)$};
        \draw[double] (7,9) -- (7,4);
        \filldraw[black] (7,5.5) circle (5pt);
        \filldraw[black] (7,6) circle (5pt);
        \filldraw[black] (7,5) circle (5pt);
        \filldraw[black] (7,4) circle (5pt);

        \draw[black] (9,10) circle (5pt);
        \draw (9,10.2) -- (9,7.8);
        \draw[black] (9,8) circle (5pt);

        \filldraw[fill=white] (10,9) rectangle (12,11) node[pos=.5]{$\mathrm{Rx}\left(\phi\right)$};
        \draw[double] (11,9) -- (11,1);
        \filldraw[black] (11,5.5) circle (5pt);
        \filldraw[black] (11,6) circle (5pt);
        \filldraw[black] (11,5) circle (5pt);
        \filldraw[black] (11,1) circle (5pt);

        \filldraw[fill=white] (13,9) rectangle (15,11) node[pos=.5]{meas};
        \draw[double] (14,9) -- (14,7);

    \end{tikzpicture}
    \caption{\textbf{Schematic of the role of classical registers throughout a qubit's lifetime.} $\left| \psi \right\rangle$ represents some other part of the graph state which the qubit may interact with. $c_1$ receives the measurement outcome of the qubit. $c_{2,3,4}$ hold the one time pad of the measurement angle, while $c_{5}$ holds the one-time-pad of the measurement outcome. Both are populated uniformly at random at the beginning of the circuit. $c_{6}$ determines if the qubit is a trap or dummy qubit and is populated depending on if this is a test round, and on a randomly chosen colour. $c_{7}$ determines if the state is initialised as $\left| + \right\rangle$ or $\left| - \right\rangle$ in the case it is a dummy and is populated uniformly at random. $c_{8}$ describes the correction required due to measurement errors and is populated depending on prior measurements of other qubits.}
    \label{fig:classical registers}
\end{figure*}

Preparing these states requires classical registers to store and update information about the state to be prepared, and to act as controls of the appropriate preparation and correction gates. To do this each vertex in the MBQC pattern graph has a corresponding eight bit classical register. We discuss here how each of these bits are populated and used, with these roles described pictorially in \cref{fig:classical registers}.

\begin{description}

    \item[Qubit measurement] The first bit of a vertex's register is used to store the results of measuring the corresponding qubit.

    \item[Measurement angle one-time pad:] The second, third and fourth bits of each vertex's classical register are used to store the one time pad of the measurement angle. That is to say to store a number from 0 to 7 used as control for the initial random rotation of each qubit, and the corresponding measurement correction.

    \item[Measurement result one-time pad:] The fifth bit is used to store the one time pad of the measurement result. That is to say a number 0 or 1 to flip the measurement outcome.

    \item[Vertex colouring:] The sixth bit is used to store if, in the case of a test round, that vertex is a dummy or a trap vertex. The sixth bit for all vertices are populated so that collectively these bits represent a colour. The user provides a complete colouring, and the particular colour is chosen at random.
    
    \item[Dummy randomness:] The seventh bit corresponds to a random choice of preparing a $\left| + \right\rangle$ or $\left| - \right\rangle$ qubit in the case of dummy vertices.

    \item[Correction tracking:] The eighth bit is used to accumulate Z corrections due to the measurement outcomes of the inverse flow of the given vertex. This bit is also used to control X corrections on neighbours of the vertex.

\end{description}

Independently of the register belonging to each vertex, there is a single bit determining if a particular shot is a compute or a test round. Additionally there are random bits used to select a random graph colour with each shot. In the experiments in \cref{sec:two qubit grover} and \cref{sec:cnot grid} a two colouring exists and so one bit is sufficient to select a random colour. In general more bits may be required for this purpose.

In addition to the registers discussed, there are a large number of scratch registers. These are used to perform calculations, such as of the corrections required based on measurement results, or to store temporary data, such as the graph colours provided by the user. Note also that qubits are measured and reset so that they may be reused. As such while each register corresponds to a graph vertex, many registers may refer to the same physical qubits at different times. There are also some corrections which take place classically, for example X corrections due to the flow, or corrections which undo the measurement result hiding.

We execute all measurement patterns in a \emph{lazy fashion}, where the qubits are initialised and entangled only when needed. Measured qubits can be reset and reused for the next sub-pattern \cite{gustiani2021blindoracle}. In this approach, the number of physical qubits required is at most $\abs{O}+1$, where $\abs{O}$ represents the number of output vertices (which here correspond to the last layer in the graph). 

In order to automate the generation of circuits implementing our on-chip verification scheme from an initial description of an MBQC computation, we present ocvqc-py \url{https://github.com/CQCL/ocvqc-py}, which we use in all of our experiments.

\subsection{Verifiable Two-Qubit Grover's Algorithm}
\label{sec:two qubit grover}

In this experiment, we verify a graph state resource in the form of a 4$\times$2 rectangular cluster, as depicted in \Cref{fig:grover1}. We implement a verifiable two-qubit Grover search algorithm, as shown in \Cref{fig:grover2}, for all possible queries. We successfully verified our execution for each query, with a low test failure rate, while consistently yielding the correct answer.

\begin{figure}[ht]
    \begin{subfigure}{\columnwidth}
        \centering
        \begin{tikzpicture}
            [baseline=(current bounding box.center),font=\footnotesize,scale=.7,auto,every node/.style={circle,inner sep=0.5pt,minimum size=8pt,draw=black,text width=9pt,align=center}]
            \node[] (n3) at (0,1){};
            \node[] (n4) at (0,0){};
            \node[] (n5) at (1,1){};
            \node[] (n6) at (1,0){};
            \node[] (n7) at (2,1){};
            \node[] (n8) at (2,0){};
            \node[] (n9) at (3,0){};
            \node[] (n10) at (3,1){};
            \draw[](n3) -- (n5);
            \draw[](n3) -- (n4);
            \draw[](n4) -- (n6);
            \draw[](n5) -- (n7);
            \draw[](n6) -- (n8);
            \draw[](n7) -- (n10);
            \draw[](n8) -- (n9);
            \draw[](n9) -- (n10);
        \end{tikzpicture}
        \caption{Verified graph state resource.}
        \label{fig:grover1}
    \end{subfigure}
    \begin{subfigure}{\columnwidth}
        \centering
        \begin{minipage}{0.59\columnwidth}
            \begin{tikzpicture}
                [baseline=(current bounding box.center),font=\footnotesize,scale=.5,auto,every node/.style={circle,inner sep=0.5pt,minimum size=6pt,draw=black,text width=7pt,align=center}]
                \node[draw=gray,fill=gray,text=white] (n3) at (0,1){2};
                \node[draw=gray,fill=gray,text=white] (n4) at (0,0){1};
                \node[draw=gray,fill=gray,text=white] (n5) at (1,1){3};
                \node[draw=gray,fill=gray,text=white] (n6) at (1,0){4};
                \node[] (n7) at (2,1){6};
                \node[] (n8) at (2,0){5};
                \node[] (n9) at (3,0){8};
                \node[] (n10) at (3,1){7};
                \draw[](n3) -- (n5);
                \draw[](n3) -- (n4);
                \draw[](n4) -- (n6);
                \draw[](n5) -- (n7);
                \draw[](n6) -- (n8);
                \draw[](n7) -- (n10);
                \draw[](n8) -- (n9);
                \draw[](n9) -- (n10);
            \end{tikzpicture}
            \hfill
            \begin{tabular}{|c|c|}
                \hline
                \footnotesize{param} & \footnotesize{angle} \\
                \hline
                $\phi_{1}$& $0$\\
                $\phi_{2}$& $0$\\
                $\phi_{5}$& $0$\\
                $\phi_{6}$& $0$\\
                $\phi_{8}$& $\pi$\\
                $\phi_{7}$& $\pi$\\
                \hline
            \end{tabular}
        \end{minipage}
        \begin{minipage}{0.39\columnwidth}
            \begin{tabular}{|c|c|}
                \hline
                \multicolumn{2}{|c|}{$\tau=0$}\\
                \hline
                $\phi_{3}$& $\pi$\\
                $\phi_{4}$& $\pi$\\
                \hline
            \end{tabular}
            \begin{tabular}{|c|c|}
                \hline
                \multicolumn{2}{|c|}{$\tau=1$}\\
                \hline
                $\phi_{3}$& $\pi$\\
                $\phi_{4}$& $0$\\
                \hline
            \end{tabular}
            
            \begin{tabular}{|c|c|}
                \hline
                \multicolumn{2}{|c|}{$\tau=2$}\\
                \hline
                $\phi_{3}$& $0$\\
                $\phi_{4}$& $\pi$\\
                \hline
            \end{tabular}
            \begin{tabular}{|c|c|}
                \hline
                \multicolumn{2}{|c|}{$\tau=3$}\\
                \hline
                $\phi_{3}$& $0$\\
                $\phi_{4}$& $0$\\
                \hline
            \end{tabular}
        \end{minipage}
        \caption{MBQC specification with a database set $\tau\in\{0,1,2,3\}$~\cite{gustiani2021blind}. The grey nodes represent the oracle operation, in which the measurement angles depend on the queried value $\tau$.}
        \label{fig:grover2}
    \end{subfigure}
    \caption{
    \textbf{Measurement based two-qubit Grover's Algorithm.}
    }
    \label{fig:grover}
\end{figure}

The measurement pattern in \Cref{fig:grover2} is used for all database queries, with the angles adjusted depending on the query. The measurement flow is linear, proceeding from the left node to the next consecutive node. We execute the measurement pattern in a lazy fashion which utilises only three physical qubits.

\begin{figure*}[t!]
    \centering
    \begin{subfigure}[t]{0.45\textwidth}
        \centering
        \includegraphics[width=\textwidth]{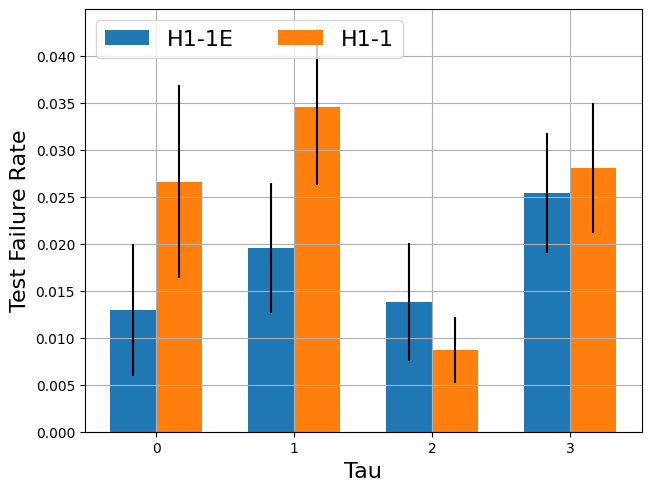}
    \end{subfigure}
    \hfill
    \begin{subfigure}[t]{0.45\textwidth}
        \centering
        \includegraphics[width=\textwidth]{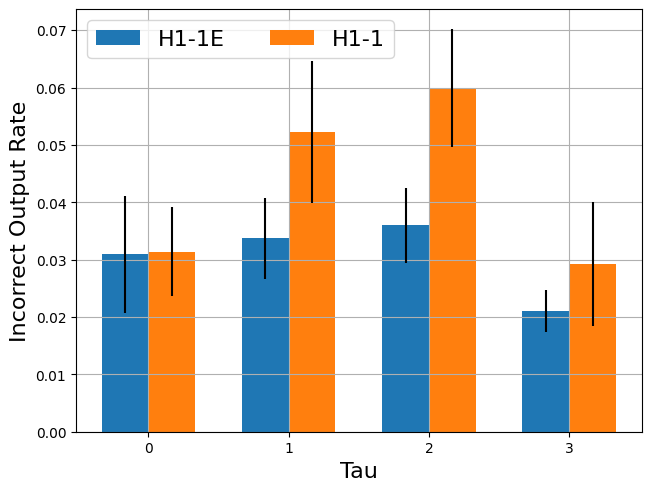}
    \end{subfigure}
    \caption{\textbf{Two qubit Grover experiment.} Results of experiment described in \cref{sec:two qubit grover}. For each value of $\tau$, we plot the test failure rate and incorrect output rate for $1000$ shots. Error bars are generated by resampling 800 shots from the data 10 times, and taking the standard deviation of the resampled test failure rate and incorrect output rate values. The bars indicate the mean of those resampled values.}
    \label{fig:two qubit grover}
\end{figure*}

The results of this experiment can be seen in \cref{fig:two qubit grover}. We see that the results from H1-1E and H1-1 are comparable, and typically within one standard deviation. However, more often than not H1-1E performs better than H1-1, indicating that the emulator underestimates the noise levels. In all cases the failure rate is comfortably less than 25\%, which is to say within the tolerance of our verification scheme. This implies that one will be able to successfully perform computations other than the Grover search on the resource of \Cref{fig:grover1} on H1-1 device, with the same confidence.

\subsection{CNOT Grid}
\label{sec:cnot grid}

\begin{figure*}
    \centering
    \begin{subfigure}[t]{0.4\textwidth}
        \centering
        \begin{tikzpicture}[thick, scale=0.4]
    
            \draw (-1,10) node[anchor=east]{$\left| b_1 \right\rangle$} -- (4,10);
            \draw (-1,9) node[anchor=east]{$\left| b_2 \right\rangle$} -- (4,9);
            \draw (-1,8) node[anchor=east]{$\left| b_3 \right\rangle$} -- (4,8);
            \draw (-1,6) node[anchor=east]{$\left| b_n \right\rangle$} -- (4,6);
    
            \draw[fill, black] (0,10) circle (5pt);
            \draw (0,10.2) -- (0,8.8);
            \draw[black] (0,9) circle (5pt);
    
            \draw[fill, black] (1,9) circle (5pt);
            \draw (1,9.2) -- (1,7.8);
            \draw[black] (1,8) circle (5pt);
    
            \draw[fill, black] (2,8) circle (5pt);
            \draw (2,8.2) -- (2,7.5);
    
            \node at (2.5,7) {$\rvdots$};
    
            \draw (3,6.5) -- (3,5.8);
            \draw[black] (3,6) circle (5pt);
    
            \node at (5,10) {$\cdots$};
            \node at (5,9) {$\cdots$};
            \node at (5,8) {$\cdots$};
            \node at (5,6) {$\cdots$};
    
            \draw (6,10) -- (11,10);
            \draw (6,9) -- (11,9);
            \draw (6,8) -- (11,8);
            \draw (6,6) -- (11,6);
    
            \draw[fill, black] (7,10) circle (5pt);
            \draw (7,10.2) -- (7,8.8);
            \draw[black] (7,9) circle (5pt);
    
            \draw[fill, black] (8,9) circle (5pt);
            \draw (8,9.2) -- (8,7.8);
            \draw[black] (8,8) circle (5pt);
    
            \draw[fill, black] (9,8) circle (5pt);
            \draw (9,8.2) -- (9,7.5);
    
            \node at (9.5,7) {$\rvdots$};
    
            \draw (10,6.5) -- (10,5.8);
            \draw[black] (10,6) circle (5pt);
    
            \draw [decorate, decoration = {calligraphic brace}] (3,5.5) --  (0,5.5) node[pos=0.5,below=5pt]{layer 1};
            \draw [decorate, decoration = {calligraphic brace}] (10,5.5) --  (7,5.5) node[pos=0.5,below=5pt]{layer m};
    
        \end{tikzpicture}
        \caption{\textbf{Circuit.} Qubits are initialised in computational basis states $b \in \left\{ 0, 1 \right\}^n$. $m$ layers of CNOT gates act on neighbouring qubits in a line architecture.}
    \end{subfigure}
    \hfill
    \begin{subfigure}[t]{0.55\textwidth}
        \centering
        \begin{tikzpicture}[thick, scale=0.4]
    
            \draw (-3,10) node[anchor=east]{$b_1 \pi$} -- (5,10);
            \draw (-3,9) node[anchor=east]{$b_2 \pi$} -- (5,9);
            \draw (-3,8) node[anchor=east]{$b_3 \pi$} -- (5,8);
            \draw (-3,6) node[anchor=east]{$b_n \pi$} -- (5,6);

            \draw[fill=white] (-3,10) circle (5pt);
            \draw[fill=white] (-3,9) circle (5pt);
            \draw[fill=white] (-3,8) circle (5pt);
            \draw[fill=white] (-3,6) circle (5pt);

            \draw[fill=white] (-2,10) circle (5pt);
            \draw[fill=white] (-2,9) circle (5pt);
            \draw[fill=white] (-2,8) circle (5pt);
            \draw[fill=white] (-2,6) circle (5pt);

            \draw[fill=white] (-1,10) circle (5pt);
    
            \draw (0,10) -- (0,9);
            \draw[fill=white] (0,10) circle (5pt);
            \draw[fill=white] (0,9) circle (5pt);

            \draw (1,9) -- (1,8);
            \draw[fill=white] (1,9) circle (5pt);
            \draw[fill=white] (1,8) circle (5pt);

            \draw (2,8) -- (2,7.5);
            \draw[fill=white] (2,8) circle (5pt);
    
            \node at (2.5,7) {$\rvdots$};
    
            \draw (3,6.5) -- (3,6);
            \draw[fill=white] (3,6) circle (5pt);

            \draw[fill=white] (4,6) circle (5pt);
    
            \node at (6,10) {$\cdots$};
            \node at (6,9) {$\cdots$};
            \node at (6,8) {$\cdots$};
            \node at (6,6) {$\cdots$};
    
            \draw (7,10) -- (15,10);
            \draw (7,9) -- (15,9);
            \draw (7,8) -- (15,8);
            \draw (7,6) -- (15,6);

            \draw[fill=white] (8,10) circle (5pt);
    
            \draw (9,10) -- (9,9);
            \draw[fill=white] (9,10) circle (5pt);
            \draw[fill=white] (9,9) circle (5pt);

            \draw (10,9) -- (10,8);
            \draw[fill=white] (10,9) circle (5pt);
            \draw[fill=white] (10,8) circle (5pt);

            \draw (11,8) -- (11,7.5);
            \draw[fill=white] (11,8) circle (5pt);
    
            \node at (11.5,7) {$\rvdots$};
    
            \draw (12,6.5) -- (12,6);
            \draw[fill=white] (12,6) circle (5pt);

            \draw[fill=white] (13,6) circle (5pt);

            \draw[fill=white] (14,10) circle (5pt);
            \draw[fill=white] (14,9) circle (5pt);
            \draw[fill=white] (14,8) circle (5pt);
            \draw[fill=white] (14,6) circle (5pt);

            \draw[fill=white] (15,10) circle (5pt);
            \draw[fill=white] (15,9) circle (5pt);
            \draw[fill=white] (15,8) circle (5pt);
            \draw[fill=white] (15,6) circle (5pt);
    
            \draw [decorate, decoration = {calligraphic brace}] (4,5.5) --  (-1,5.5) node[pos=0.5,below=5pt]{layer 1};
            \draw [decorate, decoration = {calligraphic brace}] (13,5.5) --  (8,5.5) node[pos=0.5,below=5pt]{layer m};
    
        \end{tikzpicture}
        \caption{\textbf{Measurement pattern.} Input vertices are measured by $b \in \left\{ 0, 1 \right\}^n$. All other measurements are by $0$. Measurement order is from left to right and flow is along horizontal edges. The last layer of vertices are the output.}
    \end{subfigure}
    \caption{\textbf{CNOT grid.} Equivalent circuit and MBQC model implementations of CNOT grid used in the experiments of \cref{sec:cnot grid}.}
    \label{fig:cnot grid circuit}
\end{figure*}
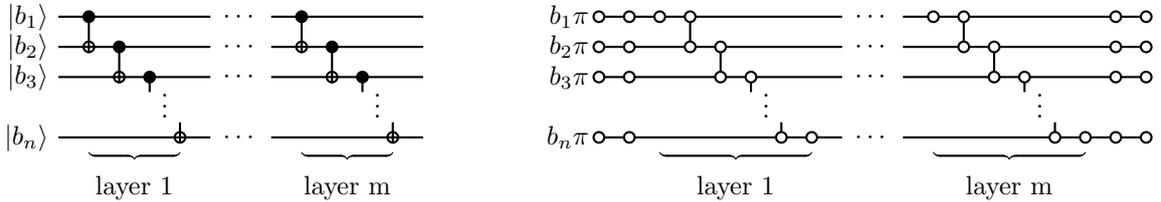

Here we implement CNOT grid circuits, as seen in \cref{fig:cnot grid circuit}, in MBQC, and verify the resulting computation. These circuits consist of $m$ layers of CNOT gates acting between $n$ neighbouring qubits in a line architecture. If the input is a tensor product of computational basis states then the ideal output can be calculated efficiently classically. As such we use this circuit as a scalable benchmark of QPUs implementing MBQC. 

As discussed in \cref{sec:verification}, one can interpret the results from the test rounds of this experiment as a benchmark of a larger class of circuits. This is namely because all computation are randomised, and so any MBQC computation with the same graph would implement indistinguishable operations. As such these CNOT grid computations give insights into the performance of computations which could not be calculated efficiently classically.

\begin{figure*}[t!]
    \centering
    \begin{subfigure}[t]{0.45\textwidth}
        \centering
        \includegraphics[width=\textwidth]{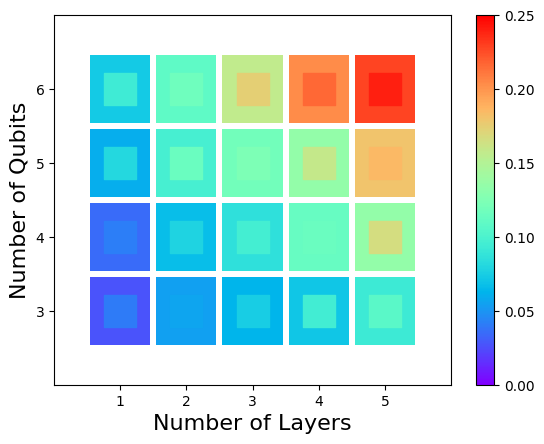}
        \caption{Colour corresponds to the test failure rate}
        \label{fig: cnot volumetric failure}
    \end{subfigure}
    \hfill
    \begin{subfigure}[t]{0.45\textwidth}
        \centering
        \includegraphics[width=\textwidth]{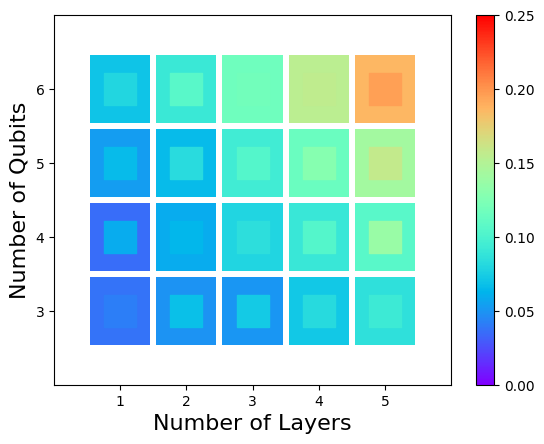}
        \caption{Colour corresponds to incorrect output rate.}
        \label{fig: cnot volumetric output}
    \end{subfigure}
    \caption{\textbf{CNOT grid H1-1E experiments.} Results of experiments described in \cref{sec:cnot grid}, run on the H1-1E device. Each outer square corresponds to the mean of 5 randomly chosen circuits of the appropriate dimension run for $1000$ shots. The inner box corresponds to the result from the worst performing circuit.}
    \label{fig: cnot volumetric}
\end{figure*}

Comprehensive results from H1-1E can be seen in the volumetric plots of \cref{fig: cnot volumetric}. Measurement patterns of increasing size have a higher test failure rate and incorrect output rate. Even in the case of the largest circuits, the correct outcome is produced ~80\% of the time. At the very largest pattern sizes we see the test failure rate reaches approximately 25\%, which is the limit of our protocol.

Note that the test failure rate grows more quickly than the incorrect final output rate. This is because the relevant output in the case of the incorrect output probability is smaller than in the case of the test failure rate, being just the last layer of qubits as opposed to roughly half the qubits which are coloured in that test round. The measurement errors the output is susceptible to is then only the corresponding output qubits, qubits which those output qubits are the flow of, or qubits whose flow neighbour the output qubits. This is an increasingly small proportion of the graph as the graph grows in size. The test failure rate, however, is susceptible to measurement errors on all qubits in the graph. 


This demonstrates that the sensitivity of our verification scheme is maintained with the size of the graph, as it captures the errors happening in the entire graph state. Thus, other computations using the same graph resource are expected to deliver the same performance. This highlights the scalability of our protocol and its robustness as a benchmark for universal computations over a predetermined graph state resource.


\begin{figure*}[t!]
    \centering
    \begin{subfigure}[t]{0.45\textwidth}
        \centering
        \includegraphics[width=\textwidth]{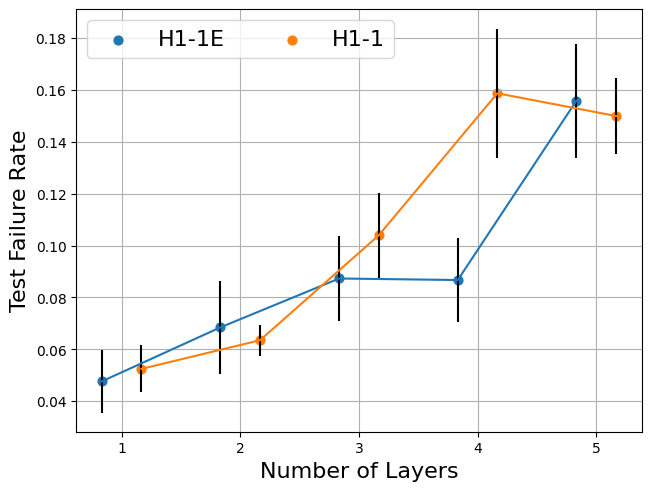}
    \end{subfigure}
    \hfill
    \begin{subfigure}[t]{0.45\textwidth}
        \centering
        \includegraphics[width=\textwidth]{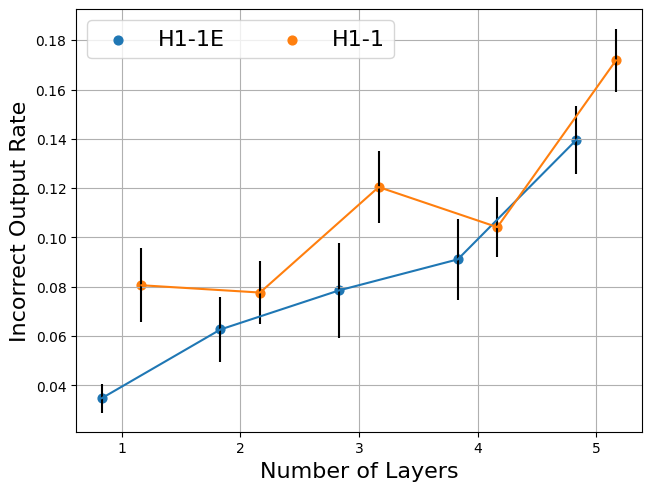}
    \end{subfigure}
    \caption{\textbf{CNOT grid H1-1 experiments.} Results of experiments described in \cref{sec:cnot grid}, run on the H1-1 device. For each number of layers, one circuit is picked at random from the circuits used in \cref{fig: cnot volumetric} and run on H1-1 for 1000 shots. Results from the same circuits on the H1-1E device are included for comparison. Error bars are generated by resampling 800 shots from the data 10 time, and taking the standard deviation of the resampled test failure rate and incorrect output rate values. The points indicate the mean of those resampled values.}
    \label{fig: machine cnot volumetric}
\end{figure*}

A subset of the circuits used in \cref{fig: cnot volumetric} were run on H1-1, with the results presented in \cref{fig: machine cnot volumetric}. At each number of layers a random circuit was selected from the five circuits used in \cref{fig: cnot volumetric} and run on H1-1. We see that H1-1 and H1-1E perform comparably, giving us confidence that the results from \cref{fig: cnot volumetric} can be expected to be indicative of the performance of H1-1 on a broader class of circuits. In all cases the average test failure rate is below 25\%. The largest measurement pattern run in \cref{fig: machine cnot volumetric} consists of 52 vertices, which to our knowledge is the largest verified MBQC computation to be run on a QPU to date.

\section{Conclusion}
\label{sec:conclusion}

In this work, we have presented an on-chip verification and benchmarking protocol for quantum computations on near-term QPUs. Our protocol requires no quantum communication between client and server, but gives guarantees on the accuracy of the computation assuming a non-adversarial server with a very general noise model. Our protocol requires only a simple tomographic experiment to ensure the assumptions of the scheme are met.

We exemplify our protocol through experimentation on the Quantinuum H1-1 device. We see strong performance on MBQC patterns with up to 52 qubits, which is, to our knowledge, the largest verified MBQC computation implemented to date. While larger graph states have been created \cite{65qubitentangle2021, Yokoyama2013}, to our knowledge no larger graph has been used for computation. Results from emulations of the H1-1 device predict strong performance with graphs containing up to 78 vertices.

In the future, we will more formally establish a benchmark for measurement-based quantum computation, using our protocol as a performance metric. As such it would also be beneficial to relate the `test failure rate' quantity used in our protocol to other more common benchmarking measures. Accordingly, we will use this benchmark to compare different quantum devices.

There exists several methods to benchmark entanglement and graph state generation \cite{entanglementdetection2020, longrangecluster2017, Zhou2019}, and graph state generation has been used as a benchmark of QPUs \cite{Wang2018, Mooney2019, 65qubitentangle2021, hamilton2022entanglementbasedvolumetricbenchmarknearterm}. However these are insufficient to benchmark a QPUs capacity to perform computation with these entangled states, as we propose to do. Benchmarks of MBQC components also exists \cite{Strydom_2023} but again do not provide a holistic view of the performance of a QPU implementing MBQC.

\paragraph{Author contributions}
C.G. performed and analysed the protocol requirements. D.L. designed the protocols and analysed their security. D.M. performed and facilitated the MBQC experiments. J.M. contributed to the execution of some experiments. R.G. provided resources and oversaw project administration. E.K. supervised the project.

\paragraph{Data Availability}
Circuits were constructed using ocvqc-py: \url{https://github.com/CQCL/ocvqc-py}. Data and scripts sufficient to recreate the plots in this paper are available: \url{https://github.com/CQCL/ocvqc-data}.

\paragraph{Acknowledgements}
We thank Alec Edgington for support with circuit construction, Vanya Eccles for support with the Nexus, and Charlie Baldwin, Cristina Cirstoiu, and Karan Khathuria for their valuable reviews of the paper.

Experiments were conducted via Quantinuum Nexus \cite{quantinuum_nexus} on the Quantinuum H1-1 QPU \cite{quantinuum_h_series}.

D.L. and E.K. acknowledge support by the Quantum Advantage Pathfinder (QAP) research programme within the UK’s National Quantum Computing Centre (NQCC).
D.L. acknowledges support by the Quantum Computing and Simulation (QCS) Hub.
D.L., J.M., and R.G. acknowledge support by the Autonomous Quantum Technologies (AutoQT) project (UKRI project 10004359).
C.G. acknowledges support by the French national quantum initiative managed by Agence Nationale de la Recherche in the framework of France 2030 with the reference ANR-22-PNCQ-0002.

\printbibliography

\appendix
\newpage 
\onecolumn
\section{Assessing Protocol
Requirements}
\label{sec:optimisation}

\subsection{Single-Qubit Tomography}
In this single qubit tomography experiment, we prepare eight quantum states $\ket{{\theta}}_{\mathit{YZ}}$, where 
\[
\ket{{\theta}}_\mathit{YZ}\equiv\ket{{\theta}} 
= \cos{\frac{\theta}{2}}\ket0 - i\sin{\frac{\theta}{2}}\ket1
\]
and 
$\theta\in\{0,\frac{\pi}{4},\frac{\pi}{2},\frac{3\pi}{4},\pi,\frac{5\pi}{4},\frac{3\pi}{2},\frac{7\pi}{4}\}$. The state $\ket{\theta}$ is prepared using two single-qubit operations:
\begin{figure}[phtb]
    \centering
    $\ketbra{{\theta}}$ = 
    \begin{quantikz}
        & \gate{\mathrm{reset}} & \gate{\mathrm{Rx}(\theta)} &
    \end{quantikz}
\end{figure}
where ``reset'' gate prepares the qubit in $\ket0$ state.
Then, for each state $\ket{\theta}$, we perform Pauli measurements in the $X$-, $Y$-, and $Z$-basis, each for 3000 shots. Therefore, we performed 24000 shots of measurements for this experiment. Finally, we construct the density matrix using the following formula: 
\[
\rho_{\theta} = 
\begin{pmatrix}
1 + r_z & r_x - i r_y \\
r_x + i r_y & 1 - r_z
\end{pmatrix},
\]
where $r_\sigma$ is the expected value of the corresponding Pauli measurement. The following table shows the  expected state preparations the tomography results.

\begin{table}[H]
    \centering
    \begin{tabular}{|c|c|c|}
        \hline 
        $\theta$  & $\ketbra{\theta}$ & $\rho_\theta$ \\ 
        \hline
        0&$\begin{pmatrix}1&0\\0&0\end{pmatrix}$&$\begin{pmatrix}1&-0.0047 - 0.005i\\-0.0047 + 0.005i&0\end{pmatrix}$\\
        $\frac{\pi}{4}$&$\begin{pmatrix}0.8536& 0.3536i\\ - 0.3536i&0.1464\end{pmatrix}$&$\begin{pmatrix}0.849&-0.0024 + 0.3581i\\-0.0024 - 0.3581i&0.151\end{pmatrix}$\\
        $\frac{\pi}{2}$&$\begin{pmatrix}0.5& 0.5i\\- 0.5i&0.5\end{pmatrix}$&$\begin{pmatrix}0.5094&0.0067 + 0.4999i\\0.0067 - 0.4999i&0.4906\end{pmatrix}$\\
        $\frac{3\pi}{4}$&$\begin{pmatrix}0.1464& 0.3536i\\ - 0.3536i&0.8536\end{pmatrix}$&$\begin{pmatrix}0.1428&-0.0077 + 0.3498i\\-0.0077 - 0.3498i&0.8572\end{pmatrix}$\\
        $\pi$&$\begin{pmatrix}0&0\\0&1\end{pmatrix}$&$\begin{pmatrix}0.0003&-0.01 + 0.0144i\\-0.01 - 0.0144i&0.9997\end{pmatrix}$\\
        $\frac{5\pi}{4}$&$\begin{pmatrix}0.1464& - 0.3536i\\0.3536i&0.8536\end{pmatrix}$&$\begin{pmatrix}0.1444&0.0193 - 0.351i\\0.0193 + 0.351i&0.8556\end{pmatrix}$\\
        $\frac{3\pi}{2}$&$\begin{pmatrix}0.5& - 0.5i\\ 0.5i&0.5\end{pmatrix}$&$\begin{pmatrix}0.498&-0.001 - 0.5i\\-0.001 + 0.5i&0.502\end{pmatrix}$\\
        $\frac{7\pi}{4}$&$\begin{pmatrix}0.8536& - 0.3536i\\ 0.3536i&0.1464\end{pmatrix}$&$\begin{pmatrix}0.8566&0.0136 - 0.3503i\\0.0136 + 0.3503i&0.1434\end{pmatrix}$ \\ \hline
    \end{tabular}
\end{table}
The diagonal terms match well, while the off-diagonal terms show some deviations. This suggests that the preparations are more affected by phase errors than flip errors, as expected.

\subsection{Convex-Optimisation}

We perform convex optimisation to calculate the bound of secret dependency, utilising the Frobenius norm $\|\cdot\|_F$ instead of the trace-norm, as defined in \Cref{eq:secret_dependent}. Specifically, we estimate a map $\mathcal E$ that minimises the distance between the ideal-world and the real-world: 
\[
\min_{\mathcal E} \frac{1}{\abs{\Theta}}\sum_{\theta\in\Theta}{\|\mathcal E (\ketbra{{\theta}})-\rho_{\theta}\|_F \coloneqq \min_{\mathcal E}\Delta(\mathcal E)},
\]
where $\Theta=\{0,\frac{\pi}{4},\frac{\pi}{2},\frac{3\pi}{4},\pi,\frac{5\pi}{4},\frac{3\pi}{2},\frac{7\pi}{4}\}$ and $\rho_\theta$ are states estimated from single-qubit tomography processes. 

To express the problem in convex optimisation, we express the parameterised map $\mathcal E$ in the form of a parameterised Choi-matrix $\Lambda_{\mathcal E}$, as the structural properties of Choi-matrices are particularly useful for numerical optimisations~\cite{wood2011tensor}, namely:
\begin{itemize}
\item Hermitian-preserving (HP): $\Lambda_{\mathcal E}^\dagger = \Lambda_{\mathcal E}$
\item Trace-preserving (TP): $\tr_B(\Lambda_{\mathcal E})=\mathbb{I}$
\item Complete-positivity (CP): $\Lambda_{\mathcal E} \geq 0$.
\end{itemize}

First, we define the parameterised Choi-matrix with 16 free real parameters $\vec x = (d_j, a_j, z_j)$:
\[
\Lambda_{\vec x} = 
\begin{pmatrix}
    d_1 & a_1 + i z_1 & a_2 + i z_2 & a_3 + i z_3 \\
    a_1-iz_1 & d_2 & a_4 + iz_4 & a_5+iz_5 \\
    a_2-iz_2 & a_4-i z_4 & d_3 & a_6 + i z_6 \\
    a_3-iz_3 & a_5-iz_5 & a_6-i z_6 & d_4
\end{pmatrix},
\]
where $0\leq d_j \leq 1$ and $\abs{a_j + z_j}\leq 1$. In this definition, $\Lambda_{\vec x}$ fulfils the HP criteria. Hence, we can define convex optimisations as the following: 
\begin{itemize}
    \item \textbf{Cost} :
    \[
    f(\vec x) = \frac{1}{8} \sum_{\theta\in \Theta} \| \Lambda_{\vec x}(\ketbra{{\theta}}) - \rho_\theta \|_F
    \]
    \item \textbf{Constrains}:
    \begin{itemize}
     \item $\tr_B(\Lambda_{\vec x}) = I $
     \item $\Lambda_{\vec x} \geq 0$,
    \end{itemize}
\end{itemize}
where $I$ is the 2-dimensional identity matrix and the Choi-matrix operation is defined as 
\[
\Lambda_{\vec x} (\rho) \coloneqq \tr_A[(\rho^T \otimes I_B)\Lambda_{\vec x}].
\]
Since the Frobenius norm is defined element-wise as $L^2$-norm, namely 
\[
\| M \|_F = \sqrt{\sum_{i,j} | M_{i,j} |^2},
\]
it is convex by nature. Finally, since our objective function $f(\vec x)$ is a sum of the Frobenius norm, hence, the convex property holds. Therefore, in this setting, we obtain the optimal solution under the Frobenius norm.

\end{document}